\newcommand{\half}{\frac{1}{2}}
\newcommand{\db}{\Omega}
\newcommand{\s}{w}

\documentclass[12pt]{article}
\usepackage{amsmath}

\date{July 23, 2013}

\numberwithin{equation}{section}

\begin{document}

\begin{title}
{The Spacetime Geometry\\ of a Null Electromagnetic Field}
\end{title}

\author{
C. G. Torre\\ {\sl Department of Physics}\\
{\it Utah State University}\\ 
{\it Logan, UT, USA, 84322-4415}\\
}
\maketitle

\begin{abstract}
We give a set of  local geometric conditions on a spacetime metric which are necessary and sufficient for it to be a null electrovacuum, that is, the metric is part of a solution to the Einstein-Maxwell equations with a null electromagnetic field.  These conditions are restrictions on a null congruence canonically constructed from the spacetime metric, and can involve up to five derivatives of the metric.  The null electrovacuum conditions are counterparts of the Rainich conditions, which geometrically characterize non-null electrovacua.  Given a spacetime satisfying the conditions for a null electrovacuum, a straightforward procedure builds the null electromagnetic field from the metric.  Null electrovacuum geometry is illustrated using some pure radiation spacetimes taken from the literature.
\end{abstract}
\bigskip\bigskip

\section{Introduction}

The purpose of this paper is to obtain local, geometric conditions on a spacetime metric  which are necessary and sufficient  for the existence of a local solution of the Einstein-Maxwell equations with everywhere null electromagnetic field.  The conditions we shall derive are counterparts of the Rainich conditions, which geometrically characterize spacetimes featuring in non-null solutions to the Einstein-Maxwell equations.   The following introduction gives the origins of this problem and summarizes the solution we have obtained. 

Let $(M, g)$ be a spacetime of signature $(-+++)$, let $\nabla$ be the torsion-free derivative compatible with $g$, let $R_{ab}$ and $R$ be the Ricci tensor and Ricci scalar of $g$, and let $F$ be a 2-form on $M$.  The Einstein-Maxwell equations (with vanishing electromagnetic sources) are
\begin{equation} 
R_{ab} - \half R g_{ab} = F_{ac}F_b{}^c - {1\over 4} g_{ab} F_{cd}F^{cd},
\label{EM1}
\end{equation}
\begin{equation}
\nabla^a F_{ab} = 0 = \nabla_{[a}F_{bc]}.
\label{EM2}
\end{equation}
Given  a solution $(g, F)$ to these equations, the underlying spacetime will be called an {\it electrovacuum} spacetime.  If the electromagnetic field $F$  satisfies 
\begin{equation}
F_{ab}F^{ab} = 0 = \epsilon^{abcd} F_{ab} F_{cd}
\label{null}
\end{equation}
at a point we say the electromagnetic field is {\it null} at that point. Here $\epsilon_{abcd}$ is the volume form determined by the metric. If a solution $(g, F)$ to (\ref{EM1}), (\ref{EM2})  satisfies (\ref{null}) everywhere we will say the spacetime is a {\it null electrovacuum}.  If the electromagnetic field nowhere satisfies (\ref{null}) we will say the spacetime is a {\it non-null electrovacuum}.

Long ago, Rainich showed that non-null electromagnetic fields can be determined purely in terms of the spacetime metric.     In particular, the following conditions on a metric $g$ are necessary and sufficient for the existence of a local solution of the Einstein-Maxwell equations with non-null electromagnetic field \cite{Rainich, MW}:
\begin{equation}
R =0, \quad R_a^b R_b^c - \frac{1}{4} \delta_a^c R_{mn}R^{mn} = 0, \quad R_{ab}t^at^b >0,
\label{AlgRain}
\end{equation}
\begin{equation}
\nabla_{[a}\left(\epsilon_{b]cde}{R^c_m\nabla^dR^{me}\over R_{ij}R^{ij}}\right) = 0.
\label{DiffRain}
\end{equation}
Here $t^a$ is any timelike vector field.  The {\it algebraic Rainich conditions} (\ref{AlgRain}) are necessary and sufficient for the existence of a  2-form $F$ which satisfies the Einstein equations (\ref{EM1}).  The {\it differential Rainich condition} (\ref{DiffRain}), which contains as many as four derivatives of the spacetime metric, is necessary and sufficient for such a 2-form to also solve the Maxwell equations (\ref{EM2}).  

The proof that the Rainich conditions are necessary and sufficient for the existence of a solution to the Einstein-Maxwell equations is constructive: there is a straightforward procedure which starts from a metric $g$ satisfying (\ref{AlgRain}),  (\ref{DiffRain}) and constructs a 2-form $F$ such that $(g, F)$ solve (\ref{EM1}), (\ref{EM2}). In fact, associated to any metric satisfying the Rainich conditions there is a 1-parameter family of electromagnetic fields.  This is because of the {\it duality symmetry} admitted by the Einstein-Maxwell equations.  If $(g, F)$ is a solution to the Einstein-Maxwell equations, then $(g, F(\theta))$ solves the Einstein-Maxwell equations, where
\begin{equation}
F(\theta) = \cos(\theta) F - \sin(\theta){}^\star F,\quad \theta \in {\bf R}.
\label{duality}
\end{equation}
Here $\star$ denotes the Hodge dual,
\begin{equation}
{}^\star F_{ab} = \frac{1}{2}\epsilon_{ab}{}^{cd}F_{cd}.
\end{equation}

Thus solutions to the Einstein-Maxwell equations with non-null electromagnetic field can be characterized in terms of ``Rainich geometry'', which is defined by a set of local geometric conditions generalizing the local condition $R_{ab}=0$ for vacuum spacetimes. Moreover, the Rainich conditions and the procedure for building the electromagnetic 2-form from a metric satisfying the Rainich conditions define an algorithm for solving the Einstein-Maxwell equations from a given input metric. This algorithm is amenable to computer automation. For example, it is implemented via the commands {\tt RainichConditions} and {\tt RainichElectromagneticField} in the Maple package {\sl DifferentialGeometry} \cite{DG}.

Evidently, the differential Rainich condition (\ref{DiffRain}) is not defined at points where $R_{ab}R^{ab}=0$.  This  occurs precisely at points where the electromagnetic field is null. Rainich geometry only describes non-null electrovacua.   The algebraic Rainich conditions (\ref{AlgRain}) are still valid in the null case, it is only the differential Rainich condition (\ref{DiffRain}) which fails.  Geroch \cite{Geroch} has extended the Rainich theory to allow for   electromagnetic fields which may be null and given a prescription for determining whether a given spacetime geometry arises from a solution of the Einstein-Maxwell equations.   Geroch's analysis thus completes  a ``geometrization'' of the electromagnetic field, insofar as it is possible. However, Geroch's analysis does not lead to a description of null electrovacua comparable to that of Rainich geometry for the non-null electrovacua.  In particular, the analysis does not provide a set of local, geometric conditions on a  metric which are necessary and sufficient for the spacetime to be a null electrovacuum.

In this paper we provide local, geometric conditions on a metric  which are necessary and sufficient  for it to be a null electrovacuum.  Like the Rainich geometry description of non-null electrovacua, the geometric characterization  of null electrovacua we obtain here is sufficiently straightforward to lead to a simple algorithm for solving the null Einstein-Maxwell equations starting from a given metric. 

It has been known for a long time that null electromagnetic fields determine a shear-free, null, geodesic congruence \cite{Mariot, Robinson}.  For this reason our analysis and results are framed within the Newman-Penrose formalism, which is especially well-suited to such situations. In the next section we set up the Einstein-Maxwell equations for a null electromagnetic field in the Newman-Penrose formalism. For the most part, this is a review of the results of  Ludwig \cite{Ludwig}, who has provided the foundation for this investigation.  In terms of the preferred congruence mentioned above, the Einstein-Maxwell equations  reduce to 3 linear inhomogeneous PDEs for one function.  The integrability conditions for these PDEs are computed and characterized in \S3 and \S4; these integrability conditions represent the principal contribution of this paper. The integrability conditions are geometric restrictions on the shear-free, geodesic, null congruence and hence they are geometric restrictions on the spacetime. They provide the local, geometric conditions for null electrovacua which replace the differential Rainich conditions (\ref{DiffRain}).    

In the generic case where the shear-free, null, geodesic congruence is twisting,  there are two (rather complicated) integrability conditions involving as many as five derivatives of the spacetime metric.  Spacetimes satisfying these conditions determine the electromagnetic field up to a duality transformation (\ref{duality}). In the special case where the congruence is twist-free, the two integrability conditions reduce to a single condition depending on as many as four derivatives of the metric and the resulting electromagnetic field depends upon an arbitrary function of one variable.\footnote{The change in the degree of uniqueness of the electromagnetic field going from  the twisting case to the  twist-free case was already recognized by Geroch and Ludwig.} In \S5 we illustrate our results using various pure radiation spacetimes taken from the literature.  In \S 6 we summarize the main results  by giving a simple algorithm which takes as input a spacetime metric and determines whether it defines a null electrovacuum and, when it does, builds the null electromagnetic field. 
 
  An Appendix provides some technical details of the analysis.  All functions considered in this paper are assumed to be $C^\infty$.  Maple worksheets which document much of this paper, provide additional examples,  and provide  code for performing various calculations can be found at

\noindent{\tt http://digitalcommons.usu.edu/dg\_applications.}  In particular, see the worksheets titled {\sl Rainich-type Conditions for Null Electrovacua}.

\section{The Einstein-Maxwell equations in the null case}

We  consider the Einstein-Maxwell  equations (\ref{EM1}), (\ref{EM2}) along with the assumption that the electromagnetic field is everywhere null, (\ref{null}).  We first review some well-known algebraic results. See, {\it e.g.}, \cite{MW}. The Einstein equations (\ref{EM1}) and the condition  (\ref{null}) imply that the Ricci tensor is null and positive, in the sense that
\begin{equation}
R_a^a = 0,\quad R_{ab}R^{bc} = 0, \quad R_{ab} t^a t^b >0,
\label{nullRicci} 
\end{equation}
where $t$ is any time-like vector field.
This is equivalent to the statement that the spacetime is a ``pure radiation'' solution of the Einstein equations. In particular, there exists a null vector field $k^a$, $g_{ab}k^a k^b=0$,  such that
\begin{equation}
R_{ab} = \frac{1}{4} k_ak_b.
\label{nullRicci2}
\end{equation}
The coefficient of $\frac{1}{4}$  in (\ref{nullRicci2}) is only for later convenience. Other normalizations could be used, although some subsequent formulas would take a different form from what we present. Notice that $k$ satisfies (\ref{nullRicci2}) if and only if $-k$ satisfies (\ref{nullRicci2}). The integral curves of the vector field $k$ determine a null congruence and it is really this congruence which is uniquely determined by the null Ricci tensor.   The contracted Bianchi identity implies that this congruence is composed of geodesics.

Thus, associated to every solution of the Einstein equations with null electromagnetic field is a null geodesic congruence $\cal C$ whose tangent vector field defines the null Ricci  (or Einstein) tensor.  Conversely,  given a metric with a  null Ricci tensor, (\ref{nullRicci2}), there exists a family of null 2-forms whose electromagnetic energy-momentum satisfies the Einstein equations. To see this, fix any spacelike unit vector field $s^a$ orthogonal to $k^a$,
\begin{equation}
g_{ab} s^a s^b = 1,\quad g_{ab} s^a k^b = 0.
\end{equation}
 Define
 \begin{equation}
 f_{ab} = k_{[a}s_{b]},
 \label{f}
 \end{equation}
 which is easily checked to be a null 2-form.
 It follows that
 \begin{equation}
 f_{ac}f_b{}^c - {1\over 4} g_{ab}f_{cd}f^{cd} = \frac{1}{4} k_ak_b.
 \end{equation}
Thus the condition (\ref{nullRicci}), equivalently (\ref{nullRicci2}), is equivalent to solving  Einstein's equation (\ref{EM1}) with a null 2-form field.  

Of course the spacelike vector field $s^a$ -- and hence $f_{ab}$  -- is far from unique.   The most general null 2-form whose energy-momentum tensor equals the Einstein tensor is of the form \cite{MW}
\begin{equation}
F_{ab} = \cos(\phi) f_{ab}  - \sin(\phi){}^\star f_{ab},
\label{generalF}
\end{equation}
where $\phi\colon M\to {\bf R}$ is any function.  This local duality transformation represents the freedom in picking the vector field $s^a$ in (\ref{f}).

To summarize: a null electrovacuum has a null Ricci tensor, which determines a null geodesic congruence $\cal C$ from which we can build the solution space (\ref{generalF}) of the Einstein equations (\ref{EM1}).  

We now wish to impose the Maxwell equations (\ref{EM2}) on the family of 2-forms (\ref{generalF}).  To this end it is advantageous to employ the Newman-Penrose formalism \cite{Stewart, Stephani}, which is defined by a choice of null tetrad.  Given the preferred congruence $\cal C$ with tangent field $k$, we define a null tetrad {\it adapted} to $k$ to be any extension of $k$ into a null tetrad. This will be any choice of null vector fields $(l, m, \overline m)$ such that the tetrad $(k, l, m, \overline m)$ satisfies
\begin{equation}
g_{ab}k^al^b = -1, \quad g_{ab}m^a \overline m^b = 1,
\end{equation}
with all other scalar products vanishing. Here $k$ and $l$ are real and $\overline m$ is the complex conjugate of $m$.  Our analysis will employ the Newman-Penrose formalism associated to any adapted tetrad. Our results will be independent of the choice of adapted tetrad and  will be determined solely by the spacetime geometry.

We fix an adapted tetrad and define
\begin{equation}
s^ a = {1\over\sqrt{2}} (m^a + \overline m^a).
\label{sdef}
\end{equation}
We substitute (\ref{generalF}) into the Maxwell equations (\ref{EM2}) and take components of the resulting equations with respect to the adapted tetrad. We then express everything in terms of the Newman-Penrose spin coefficients, as defined in \cite{Stewart, Stephani}, and directional derivatives, defined by
\begin{equation}
D = k^a\nabla_a,\quad \Delta = l^a\nabla_a, \quad \delta = m^a\nabla_a,\quad \overline\delta = \overline m^a\nabla_a.
\end{equation}
The resulting equations are \cite{Ludwig}:
\begin{equation}
\kappa = 0 = \sigma
\label{SFNG}
\end{equation}
\begin{equation}
{1\over i} D\phi - 2\epsilon + \rho =  0, \quad {1\over i} \delta \phi + \tau - 2\beta = 0,\quad i\overline\delta \phi+ \overline\tau - 2\overline \beta = 0.
\label{phieq}
\end{equation}
Thus a null electrovacuum determines a shear-free ($\sigma=0$),  geodesic ($\kappa = 0$), null congruence and, given an adapted tetrad, a function $\phi$ satisfying 3 linear, inhomogeneous PDEs.

\bigskip\noindent
{\sl Remarks:}

\begin{itemize}

\item Another way to obtain (\ref{SFNG}) and (\ref{phieq}) is to start with the Newman-Penrose form of the Maxwell equations (see \cite{Stewart, Stephani}).  In an adapted tetrad, the definitions (\ref{f}) and (\ref{generalF}) imply the electromagnetic scalars satisfy $\Phi_0 = \Phi_1=0$, $\Phi_2 = \frac{1}{4}e^{i\phi}$.   The Maxwell equations immediately reduce to (\ref{SFNG}) and (\ref{phieq}).¡

\item From the contracted Bianchi identity and  the normalization (\ref{nullRicci2}) it follows that the parametrization of the geodesics defined by $k$ is such that in an adapted tetrad we  have
\begin{equation}
\Re(\rho) = 2\Re(\epsilon).
\label{parm}
\end{equation}
This is the real part of the first equation in (\ref{phieq}), so  (\ref{phieq})  represents 3 real PDEs for $\phi$.  Notice that the geodesics determined by $k$ are affinely parametrized only if the divergence of $k$ vanishes, that is, $\Re(\rho) = 2\Re(\epsilon)=0$.

\item From (\ref{SFNG}), (\ref{nullRicci2}), and  the Goldberg-Sachs theorem \cite{Stephani}, all null electrovacua are algebraically special with $k$ being a repeated principal null direction.  

\item Using the adapted null tetrad, the Newman-Penrose curvature scalars satisfy:
\end{itemize}

\begin{align}
\Psi_0 = \Psi_1 =\Phi_{00} = \Phi_{01} = \Phi_{10} =  \Phi_{02} = \Phi_{20} =\Phi_{11}= \Phi_{12} = \Phi_{21} = \Lambda =  0 .
\label{NPCurvature}
\end{align}

The Maxwell equations for null electromagnetic fields, (\ref{SFNG}) and (\ref{phieq}), are constructed from a tetrad adapted to $k$. However these equations do not depend in any essential way upon the choice of $k$ or tetrad adapted to $k$. To see this we proceed as follows. First, in general any two null tetrads differ by a transformation from the group of ``local  Lorentz transformations'', which is the group of mappings from $M$ into the Lorentz group.  Any two tetrads adapted to $k$ will then be related by a local Lorentz transformation which fixes $k$. Any such transformation is a composition of 2 types of local Lorentz transformations:  a local spatial rotation and a local null rotation about $k$. These transformations are given, respectively, by
\begin{equation}
(k, l, m, \overline m) \longrightarrow (k, l, e^{2i \theta} m, e^{-2i\theta} \overline m),
\label{SR}
\end{equation}
\begin{equation}
(k, l, m, \overline m)  \longrightarrow (k, l + cm + \overline c\, \overline m + |c|^2 k, m + \overline c k, \overline m + c k),
\label{NR}
\end{equation}
where $\theta\colon M\to {\bf R}$ and $c\colon M\to {\bf C}$.  These changes of tetrad induce well-known  transformation laws for the Newman-Penrose spin coefficients; see {\it e.g.,} \cite{Stewart}. From these transformation laws it follows that ({\ref{SFNG}) and (\ref{parm}) are preserved under a change of adapted tetrad. Moreover, if $\phi$ solves (\ref{phieq}) for one choice of adapted tetrad, then under a change of tetrad, as described above, $\phi - 2\theta$ satisfies (\ref{phieq}) for the transformed tetrad \cite{Ludwig}.  In this sense the group of local Lorentz transformations mapping one adapted tetrad to another is a symmetry group for the Maxwell equations for null fields, and equations (\ref{SFNG}), (\ref{phieq}) are independent of the choice of tetrad adapted to $k$.   

Finally, it is straightforward to verify that   the transformation 
\begin{equation}
k \to - k,\quad  l \to - l
\label{redef}
\end{equation} 
is also a symmetry  of the equations (\ref{SFNG}), (\ref{phieq}), and (\ref{parm}).  More precisely, these equations transform homogeneously at most.  Therefore, if any equation from (\ref{SFNG}), (\ref{phieq}), and (\ref{parm}) is satisfied it will still be satisfied after making the redefinition (\ref{redef}). This transformation is relevant since the spacetime geometry only determines the vector field $k$ up to a sign via (\ref{nullRicci2}).  This symmetry and the local Lorentz symmetry described above guarantee that the equations   (\ref{SFNG}), (\ref{phieq}), and (\ref{parm}) are geometrically defined by the spacetime metric only.   We will have  more to say on this point in \S4.

\section{Integrability Conditions}

Let us summarize the situation so far.  Let $g$ define a null electrovacuum. It follows that the Ricci tensor of $g$ is null.  The null Ricci tensor determines a null congruence in spacetime which must be geodesic and shear-free.  In terms of this congruence the Einstein-Maxwell equations reduce to 3 linear, inhomogeneous PDEs (\ref{phieq}) for a function $\phi$.  We now compute the integrability conditions for these PDEs. These conditions, which are additional geometric conditions on the congruence, will lead to the null electrovacuum counterpart of the differential Rainich condition (\ref{DiffRain}) for non-null electrovacua.  

The nature of the integrability conditions for (\ref{phieq}) depends upon whether the distribution $\cal D$, defined as the span of $(k, m, \overline m)$ at each point of $M$, is integrable, that is, whether  $[{\cal D}, {\cal D}] \subset {\cal D}$.  In terms of the Newman-Penrose formalism for any null tetrad $(k, l, m, \overline m)$  we have \cite{Stewart}:

\begin{align}
&[m, k] = (\overline\alpha + \beta - \overline \pi)k + \kappa l - (\overline \rho + \epsilon - \overline\epsilon) m - \sigma\overline m,\nonumber\\
&[\overline m, k] = (\alpha + \overline\beta -  \pi)k +\overline\kappa l -\overline\sigma m - ( \rho + \overline\epsilon - \epsilon)\overline m ,\nonumber\\
&[\overline m, m] = (\overline\mu - \mu) k +(\overline\rho - \rho) l + (\alpha -\overline\beta) m - (\overline\alpha -\beta)\overline  m,\nonumber\\
&[m, l] = - \overline\nu k + (\tau -\overline\alpha - \beta) l + (\mu - \gamma + \overline\gamma)m + \overline\lambda\, \overline m\nonumber\\
&[\overline m, l] = - \nu k + (\overline\tau -\alpha - \overline\beta) l + \lambda\, m + (\overline\mu - \overline\gamma + \gamma)\overline m \nonumber\\
&[k, l] = -(\gamma + \overline\gamma) k - (\epsilon + \overline\epsilon) l + (\overline\tau + \pi) m + (\tau +\overline\pi)\overline m .\nonumber\\
\label{brackets}
\end{align}
Evidently, given $\kappa=0$, $\cal D$ is integrable if and only if the rotation -- or ``twist'' -- scalar $\omega$ vanishes, where
\begin{equation}
\omega = - \Im(\rho).
\end{equation}
The distribution $\cal D$ and its integrability is a property of the null geodesic congruence $\cal C$ and is independent of the choice of $k$ or tetrad adapted to it.   

In what follows we will also need the dual basis to the null tetrad and its structure equations.  We denote the basis of 1-forms dual to $e_\alpha = (k, l, m, \overline m)$ by $\db^\alpha = (\db_k, \db_l, \db_m, \db_{\overline m})$.  The dual basis satisfies
\begin{equation}
\db^\alpha(e_\beta) = \delta^\alpha_\beta
\end{equation}
and
\begin{align}
d\db_k = &(\gamma +\overline\gamma) \db_k \wedge \db_l + (\overline\alpha + \beta - \overline \pi) \db_k\wedge\db_m+ (\alpha + \overline\beta - \pi)\db_k \wedge \db_{\overline m}\nonumber\\
& - \overline \nu \db_l \wedge\db_{\overline m}
- (\mu - \overline\mu)\db_ m \wedge \db_{\overline m}
\label{structure1}
\end{align}
\begin{align}
d\db_l = &(\epsilon + \overline\epsilon)\db_k \wedge \db_l +\kappa \db_k \wedge \db_ m  + \overline\kappa \db_k \wedge \db_{\overline m}
 + (\tau -\overline\alpha - \beta) \db_l \wedge \db_m\nonumber\\
 & + (\overline\tau -\alpha - \overline\beta) \db_l \wedge \db_{\overline m} - (\rho - \overline\rho) \db_m \wedge \db_{\overline m}
 \label{structure2}
 \end{align}
 \begin{align}
 d\db_m = &(\overline \tau + \pi) \db_l \wedge \db_k + (\overline \epsilon - \epsilon - \overline\rho) \db_k \wedge\db_m + (\mu - \gamma +\overline\gamma)\db_l \wedge \db_m\nonumber\\
  &+ (\alpha - \overline\beta) \db_m \wedge \db_{\overline m} - \overline\sigma \db_k \wedge \db_{\overline m} + \lambda \db_l \wedge \db_{\overline m}
  \label{structure3}
 \end{align}
  \begin{align}
 d\db_{\overline m} = &( \tau + \overline\pi) \db_l \wedge \db_k + ( \epsilon -\overline \epsilon - \rho) \db_k \wedge\db_{\overline m} + (\overline\mu - \overline\gamma +\gamma)\db_l \wedge \db_{\overline m}\nonumber\\
  &- (\overline\alpha - \beta) \db_m \wedge \db_{\overline m} - \sigma \db_k \wedge \db_{ m} + \overline\lambda \db_l \wedge \db_{ m}
  \label{structure4}
 \end{align}
\subsection{Twisting case, $\omega\neq0$}

In this sub-section we assume that $\omega \neq0$ and  that (\ref{SFNG}) and (\ref{parm}) hold.  Thus the null congruence $\cal C$ is twisting, shear-free, and geodesic. Consider the identity
\begin{equation}
[\delta,\overline\delta]\phi - \delta(\overline\delta\phi) + \overline\delta(\delta\phi) = 0.
\end{equation}  
Using (\ref{brackets})  to compute  the bracket and using (\ref{phieq}) to evaluate the various derivatives, we obtain an additional differential equation for $\phi$:
\begin{equation}
\omega\Delta\phi   + \half (\mu - \overline\mu)(\rho - 2\epsilon) + \Re\left[(\overline\delta+\overline\beta-\alpha)(\tau-2\beta)  \right]=0.
\label{ic0}
\end{equation}
Thus the equations determining $\phi$ are (\ref{phieq}) and their consequence (\ref{ic0}). These equations can be written in terms of differential forms as
\begin{equation}
d\phi = A,
\end{equation}
where $A$ is a (real) 1-form given by
\begin{align}
A =   &i(2\epsilon - \rho) \db_k - {1\over\omega}\left\{ \half (\mu - \overline\mu)(\rho - 2\epsilon) + \Re\left[(\overline\delta+\overline\beta-\alpha)(\tau-2\beta)  \right]\right\}\db_l\nonumber\\
&+ i(2\beta - \tau)\db_m - i(2\overline\beta - \overline\tau)\db_{\overline m}.
\end{align}
Therefore, according to the Poincar\'e lemma \cite{Lang}, local smooth solutions $\phi$ exist, unique up to addition of a constant, if and only if
\begin{equation}
dA = 0.
\end{equation}
This condition represents 6 equations which can be obtained by contraction with pairs of basis vectors. These equations are
\begin{equation}
{\cal I}_1 = {\cal I}_2 = {\cal I}_3 = {\cal I}_4 = {\cal I}_5 = {\cal I}_6 = 0,
\end{equation}
where, using (\ref{structure1})--(\ref{structure4}), we have
\begin{align}
{\cal I}_1 &\equiv - \omega^2 dA(m, l)\nonumber\\
&={\omega}\delta \Big\{\Re\Big[\delta(\overline\tau-2\overline\beta)\Big]\Big\} - \Big[\delta \omega + \omega(\tau - \overline\alpha - \beta)\Big]\Big[ \Re\left\{(\overline\delta + \overline\beta - \alpha)(\tau - 2\beta)\right\}\nonumber\\
&+ i\Im(\mu)(\rho - 2\epsilon)\Big]
+\frac{\omega}{2}\Big\{\overline\beta\delta(2\overline\alpha + \tau - 4\beta) + \beta\delta(2\alpha + \overline\tau - 4\overline\beta) + 2i\delta(\Im(\mu)(\rho-2\epsilon))\nonumber\\ 
&+ \tau\delta(\overline\beta - \alpha) +  \overline\tau\delta(\beta - \overline\alpha)
- \alpha\delta(\tau - 2\beta) - \overline\alpha\delta(\overline\tau - 2\overline\beta)\Big\}
- i\omega^2\Delta(\tau - 2\beta)\nonumber\\
&+\omega^2\Big[\overline\nu(\omega + 2\Im(\epsilon)) -(\tau-2\beta)(2\Im(\gamma) + i\mu)  +i\overline\lambda (\overline\tau - 2\overline\beta) \Big],
\label{IC1}
\end{align}
\begin{align}
{\cal I}_2 &\equiv - \omega^2 dA(\overline m, l)\nonumber\\
&={\omega}\overline\delta \Big\{ \Re\Big[\delta(\overline\tau-2\overline\beta)\Big]\Big\}
- \Big[\overline\delta \omega + \omega(\overline\tau - \alpha - \overline\beta)\Big]\Big[ \Re\left\{(\overline\delta + \overline\beta - \alpha)(\tau - 2\beta)\right\}\nonumber\\ 
&-i\Im(\mu)(\overline\rho - 2\overline\epsilon)\Big]
+\frac{\omega}{2}\Big\{\beta\overline\delta(2\alpha + \overline\tau - 4\overline\beta) + \overline\beta\, \overline\delta(2\overline\alpha + \tau - 4\beta)- 2i\overline\delta(\Im(\mu)(\overline\rho-2\overline\epsilon))\nonumber\\ 
&+ \overline\tau\overline\delta(\beta - \overline\alpha) +  \tau\overline\delta(\overline\beta - \alpha)
- \overline\alpha\overline\delta(\overline\tau - 2\overline\beta) - \alpha\overline\delta(\tau - 2\beta)
\Big\} +i\omega^2\Delta(\overline\tau - 2\overline\beta)\nonumber\\
&+\omega^2\Big[\nu(\omega + 2\Im(\epsilon)) -(\overline\tau-2\overline\beta)(2\Im(\gamma) - i\overline\mu)  -i\lambda (\tau - 2\beta)\Big],
\label{IC2}
\end{align}
\begin{align}
{\cal I}_3 &\equiv - \omega^2 dA(l,k)\nonumber\\
&=\omega^2\Big\{-2 \Re(\gamma) \left[\omega + 2\Im(\epsilon)\right]
-2\Im\left[(\overline\tau +\pi)(\tau - 2\beta) + i\Delta(\rho-2\epsilon)\right]\Big\} \nonumber\\
& + \left[\Re\left((\overline\delta - \alpha +\overline\beta)(\tau - 2\beta)\right) + {i\over4} (\rho-2\epsilon)\Im(\mu)
- {3\over 4} i \Im(\mu)(\overline\rho - 2\overline\epsilon)\right]\left[-{2\omega}\Re(\epsilon) + D\omega\right] \nonumber\\
&-\omega \Big\{-\Re((\tau-2\beta)D\alpha) + \tau\Re(D\beta) + i\rho\Im(D\mu) + \Im(\mu) D\omega  + 2\Im(D\epsilon)\Im(\mu) \nonumber\\
&+2\Re\left[\half D\overline\delta (\tau - 2\beta) - 2\overline\beta D\beta - \epsilon D\mu\right]
- i \Im(\tau)D\beta + \Re[2\alpha D\beta + (\overline\beta - \alpha)D\tau]\nonumber\\
 &+ 2\Re(\epsilon)D\overline\mu
\Big\},\label{IC3}
\end{align}

\begin{align}
{\cal I}_4 &\equiv -idA(m, k)\nonumber\\
&=(D-\overline\rho)(\tau - 2\beta) + (\delta -\overline\alpha +\beta +\overline\pi - \tau) (\epsilon - \overline\epsilon) +i(\delta +\overline\pi -\overline\alpha - \beta)\omega,
\label{trivial1}
\end{align}

\begin{align}
{\cal I}_5 &\equiv idA(\overline m, k)\nonumber\\
&=
(D-\rho)(\overline\tau - 2\overline\beta) - (\overline\delta -\alpha +\overline\beta +\pi - \overline\tau) (\epsilon - \overline\epsilon) -i(\overline\delta +\pi -\alpha - \overline\beta)\omega,
\label{trivial2}
\end{align}

\begin{equation}
{\cal I}_6 \equiv dA(m, \overline m) = 0.
\end{equation}
Evidently, condition ${\cal I}_6 = 0$ is trivial. We shall see in \S4 that conditions ${\cal I}_3 = 0$, ${\cal I}_4=0$ and ${\cal I}_5=0$ are also trivial, so that it is only  the conditions $ {\cal I}_1 = {\cal I}_2 = 0$ on the twisting, shear-free, null, geodesic congruence $\cal C$ which define null electrovacua.

\subsection{Twist-free case, $\omega=0$}

Here we consider the special case where the congruence has everywhere vanishing twist, $\omega=0$.  We still assume that (\ref{SFNG}) and (\ref{parm}) hold.  In this case the distribution ${\cal D}\subset TM$ defined by 
\begin{equation}
{\cal D} = \bigcup_{p\in M} {\cal D}_p \, ,
\end{equation}
\begin{equation}
{\cal D}_p = {\rm span}\{k, m, \overline m\}\big|_p \, 
\end{equation}
 is integrable.  We can then write $M = {\bf R} \times \Sigma$ where we label the embedded null hypersurfaces $\Sigma$ by $u\in {\bf R}$. For each $u$,  the vector fields $(k, m, \overline m)$ are everywhere tangent to  $\Sigma$.  This foliation of $M$ by $\Sigma$ is determined by the congruence $\cal C$ alone and is independent of the choice of tetrad adapted to $k$.  We denote by $(\tilde \db_k, \tilde \db_m, \tilde \db_{\overline m})$ the basis dual to  $(k, m, \overline m)$ on $\Sigma$ for each $u$.
  
The equations (\ref{phieq}) can now be viewed as a 1-parameter family of  differential equations  for a 1-parameter family of functions $\tilde\phi(u)\colon \Sigma \to {\bf R}$.   With this interpretation we can express these equations as
\begin{equation}
d\tilde\phi(u) = \tilde A(u),\quad {\rm on\ \ } \Sigma,
\end{equation}
where $\tilde A(u)$ is a 1-parameter family of 1-forms on $\Sigma$ given by
\begin{align}
\tilde A(u) =   i(2\epsilon(u) - \rho(u))\tilde  \db_k 
+ i(2\beta(u) - \tau(u))\tilde \db_m - i(2\overline\beta(u) - \overline\tau(u))\tilde \db_{\overline m}.
\end{align}
Therefore, according to the Poincar\'e lemma, local smooth solutions $\tilde \phi(u)\colon \Sigma\to {\bf R}$ exist, unique up to an additive constant, if and only if 
\begin{equation}
 d \tilde A(u) = 0\quad {\rm on\ } \Sigma.
\label{ICTF}
\end{equation}
Provided (\ref{ICTF}) holds for all $u$, it follows from the form of the solution provided by the homotopy operator proof of the Poincar\'e lemma that the solutions  $\tilde \phi(u)$ determine smooth local solutions  $\phi\colon M\to {\bf R}$ of (\ref{phieq}). The resulting solution $\phi$ will be unique up to addition of a function of $u$.

The integrability condition $ d \tilde A = 0$ represents 3 equations  which can be obtained by contraction with pairs of  vectors taken from $(k, m, \overline m)$. These equations are (for each value of $u$ --- we leave this variable implicit for simplicity)
\begin{equation}
 {\cal J}_1 = {\cal J}_2 = {\cal J}_3 = 0,
 \end{equation}
 where
 \begin{equation} 
{\cal J}_1 \equiv -{i\over 2} d\tilde A(m, \overline m)=
\Re\left[(\overline\delta+\overline\beta-\alpha)(\tau-2\beta)\right] - \half (\mu - \overline\mu)(\epsilon-\overline\epsilon),
\label{IC4}  
\end{equation}
\begin{equation}
{\cal J}_2 \equiv -id\tilde A(m, k) = {\cal I}_4\big|_{\omega=0},\quad {\cal J}_3\equiv id\tilde A(\overline m, k)
 = {\cal I}_5\big|_{\omega=0}.
 \end{equation}
In the next section we shall see that the conditions ${\cal J}_2 = 0 = {\cal J}_3$ are trivial, so that it is only the condition ${\cal J}_1 = 0$ on the twist-free, shear-free,  null, geodesic congruence $\cal C$ which defines null electrovacua. 

Finally, we note that the  integrability conditions for the twisting case, as defined in the previous section, become trivial in the limit $\omega \to0$.  In this limit one of the equations for $\phi$,  (\ref{ic0}), reduces to the integrability condition ${\cal J}_1=0$ for the remaining equations (\ref{phieq}).  In this way a null electrovacuum with a twisting congruence reduces to a null electrovacuum with a twist-free congruence.

\section{Invariance Properties and a Minimal Set of  Integrability Conditions}

 We now consider in some detail the transformation of the integrability conditions induced by a change of adapted tetrad, as described in \S2.  The first result we need is the transformation properties of $A$ and $\tilde A$ under a change in tetrad involving a local spatial rotation (\ref{SR}), a local null rotation  (\ref{NR})  and a change in the sign of $k$.     For all computations in this section we will assume the null congruence $\cal C$ is geodesic and shear-free, (\ref{SFNG}), and that (\ref{parm}) holds. 

The transformation of  spin coefficients under such transformations of the null tetrad can be found in many standard references, see, {\it e.g.}, \cite{Stewart}.  The transformations of the dual basis on $M$ induced by the transformations (\ref{SR}), (\ref{NR}), and $k \to -k$ are given by, respectively,
\begin{align}
(\Omega_k, \Omega_l, \Omega_m,\Omega_{\overline m}) &\quad\to\quad (\Omega_k, \Omega_l, e^{-2i\theta}\Omega_m,e^{2i\theta}\Omega_{\overline m}),
\nonumber\\
 &\quad\to\quad (\Omega_k - \overline c\, \Omega_m - c\, \Omega_{\overline m} + |c|^2\,  \Omega_l, \Omega_l, \Omega_m - c\,  \Omega_l,\Omega_{\overline m} - \overline c\, \Omega_l),\nonumber\\
 &\quad\to\quad (-\Omega_k, -\Omega_l, \Omega_m,\Omega_{\overline m})
\end{align}
The transformations of the dual basis $(\tilde \Omega_k, \tilde  \Omega_m, \tilde \Omega_{\overline m})$ to $(k, m, \overline m)$ on $\Sigma$ are given by
\begin{align}
(\tilde \Omega_k, \tilde \Omega_m,\tilde \Omega_{\overline m}) &\quad\to\quad (\tilde \Omega_k,  e^{-2i\tilde\theta}\tilde \Omega_m,e^{2i\tilde\theta}\tilde \Omega_{\overline m}),
\nonumber\\
 &\quad\to\quad (\tilde \Omega_k - \overline{\tilde c}\, \tilde  \Omega_m - \tilde c\,  \tilde \Omega_{\overline m} ,  \tilde \Omega_m,\tilde \Omega_{\overline m} ),\nonumber\\
&\quad\to\quad (-\tilde \Omega_k,\tilde  \Omega_m,\tilde \Omega_{\overline m})
\end{align}
Here we are viewing $\tilde c$ and $\tilde \theta$ as 1-parameter families of functions on $\Sigma$.

We now use the transformation of the spin coefficients and the change of dual basis shown above to compute the change in $A$ and $\tilde A$. A straightforward computation   reveals that the 1-forms $A$ on $M$ and  $\tilde A$ on $\Sigma$  are unchanged by the redefinition  $k \to - k$.  A more lengthy but straightforward computation shows that $A$ and $\tilde A$ are unchanged by a local null rotation.\footnote{The Newman-Penrose Ricci identities are come into play when checking the invariance of $A$ under null rotations.}  Finally, under a local spatial rotation we have
\begin{align}
&A \to A - 2d\theta,\quad {\rm on\ } M\\
&\tilde A \to \tilde A - 2d\tilde \theta,\quad {\rm on\ } \Sigma.
\end{align}
It  follows that the 2-forms  $dA$ on $M$ and $d\tilde A $ on $\Sigma$ are  unchanged by a change of adapted tetrad and/or a change in sign of $k$.  Thus the integrability conditions $dA=0$ on $M$ and $d\tilde A = 0$ on $\Sigma$ represent geometric restrictions on $\cal C$, which can be viewed as local, geometric conditions on the spacetime metric. 

 Using the invariance of $dA$ under the transformations (\ref{SR}), (\ref{NR}), and $k \to -k$ we easily infer from their definitions that ${\cal I}_4$ and ${\cal I}_5$  transform by simple rescalings, at most.  In addition, ${\cal I}_3$ transforms into a linear combination of itself and ${\cal I}_4$ and ${\cal I}_5$. Thus if  
 ${\cal I}_3 = {\cal I}_4={\cal I}_5=0$ for any one tetrad adapted to any choice of $k$, it will hold for all tetrads adapted to any choice of $k$.  Similarly,  ${\cal J}_2$ and ${\cal J}_3$ transform by simple rescalings at most. Thus, if ${\cal J}_2 ={\cal J}_3 = 0$ hold for any one tetrad adapted to any choice of $k$, they will hold for all tetrads adapted to any choice of $k$.  
These observations allow for a relatively quick proof of the fact that {\it the conditions ${\cal I}_3 = {\cal I}_4={\cal I}_5=0$ in the twisting case, and ${\cal J}_2 =  {\cal J}_3=0$ in the twist-free case, are all trivially satisfied for the shear-free null geodesic congruence $\cal C$ determined by the null Ricci tensor.}  The proof involves using a  convenient adapted tetrad and then using some of the Newman-Penrose Ricci identities; it is given in the Appendix.  

Thus the only conditions to be considered are ${\cal I}_1 = {\cal I}_2=0$  in the twisting case and ${\cal J}_1=0$ in the twist-free case.  Because ${\cal I}_3 = {\cal I}_4={\cal I}_5={\cal I}_6=0$ and ${\cal J}_2 =  {\cal J}_3=0$, it follows that the conditions ${\cal I}_1 = {\cal I}_2=0$ and  ${\cal J}_1=0$ are invariant under  local null rotations, under spatial rotations,  and under a change of sign of $k$. The   invariants ${\cal I}_1$ and ${\cal I}_2$ are non-trivial. This follows from the  example in \S5.4, where these invariants are shown to be non-zero for a particular class of twisting, pure radiation spacetimes.  The   invariant ${\cal J}_1$ is non-trivial. This follows from the  example in \S5.2.2, where it is shown to be non-zero for a particular class of twist-free, pure radiation spacetimes. 

To summarize:  In the twisting case, the integrability conditions are
\begin{equation}
{\cal I}_1 = {\cal I}_2=0,
\label{twistIC}
\end{equation} 
where ${\cal I}_1$ and ${\cal I}_2$ are defined in (\ref{IC1}) and (\ref{IC2}).
This set of conditions is invariant with respect to spatial rotations, null rotations, and change of sign of $k$. Thus these conditions on the twisting, shear-free, null geodesic congruence  $\cal C$ represent geometric conditions on the spacetime metric.  In the twist-free case, the integrability condition is 
\begin{equation}
{\cal J}_1=0,
\label{twistfreeIC}
\end{equation} 
where ${\cal J}_1$ is defined in (\ref{IC4}). This condition is unchanged by spatial rotations, null rotations, and change of sign of $k$. Thus this condition on the twist-free, shear-free, null geodesic congruence $\cal C$ represents a geometric condition on the spacetime metric.

\section{Examples}

We use the results of \S3 and \S4 to study four classes of pure radiation spacetimes taken from the literature on solutions of the Einstein field equations.

\subsection{A homogeneous pure radiation spacetime}

Here we consider a one parameter family of spacetimes which, until recently, was believed to represent the only homogeneous, pure radiation solutions of the Einstein equations which are {\it not}  null electrovacua \cite{Stephani}. However, in \cite{CGT} it was shown that these spacetimes are in fact null electrovacua by exhibiting the electromagnetic sources for the spacetimes. The results of \cite{CGT} were originally obtained from the analysis of the previous sections, which we now demonstrate.

In terms of coordinates $(u, v, x, y)$ on ${\bf R}^4$ and a parameter $\s$, the homogeneous, pure radiation metric in question  is given by
\begin{equation}
g := -2 e^{2\s x} du \otimes du + du \otimes dv + dv \otimes du + dx \otimes dx + dy \otimes dy.
\label{ex1metric}
\end{equation}
(See Theorem 12.6 in \cite{Stephani}.) The Ricci tensor ${\cal R}$ of this metric is given by
\begin{equation}
{\cal R} = 4\s^2 e^{2\s x} du \otimes du,
\end{equation}
which is readily seen to be null, satisfying (\ref{nullRicci}). The null vector field $k$ defined in (\ref{nullRicci2}), tangent to the null congruence $\cal C$, is given by
\begin{equation}
k = 4\s e^{\s x}\partial_v.
\end{equation}
A null tetrad $(k, l, m, \overline m)$ adapted to $k$ is
\begin{align}
&k = 4\s e^{\s x}\partial_v,\quad l = -\frac{1}{4\s}(e^{-\s x}\partial_u + e^{\s x}\partial_v),\nonumber\\
& m = \frac{1}{\sqrt{2}}(\partial_x + i \partial_y),\quad \overline m = \frac{1}{\sqrt{2}}(\partial_x - i \partial_y).
\end{align}
The Newman-Penrose spin coefficients for this tetrad are
\begin{equation}
\alpha = \beta=\frac{\sqrt{2}}{4} \s,\quad \nu = \frac{\sqrt{2}}{16\s},\quad \epsilon =\gamma = \kappa=\lambda=\mu=\pi=\rho=\sigma=\tau = 0.
\label{ex1Spin}
\end{equation}
From $\kappa=0 = \sigma$ we see that the null congruence is geodesic and shear-free. From $\rho=0$ we see that the twist  of the congruence vanishes. The  associated foliation of spacetime is by the null hypersurfaces $u=const.$  Because the congruence is twist-free, the only remaining condition for a null electrovacuum is (\ref{twistfreeIC}), which is easily seen to be satisfied by the spin coefficients in (\ref{ex1Spin}).  Thus the metric (\ref{ex1metric}) is a null electrovacuum.  

We now use equations (\ref{phieq})  to compute the electromagnetic field.   After some minor simplifications,  these three equations respectively take the  form
\begin{equation}
-i \partial_x \phi + \partial_y \phi - \s = 0,\quad
i\partial_x\phi + \partial_y\phi -\s=0,\quad
\partial_v\phi = 0.
\end{equation}
The general solution to these equations is
\begin{equation}
\phi = \s y + h(u),
\end{equation}
where $h$ is an arbitrary function.
Choosing $s = \partial_x$ in (\ref{f}) and then constructing $F$ via (\ref{generalF}) we get the null 2-form
\begin{equation}
F = 2\s \cos(\s y+ h(u))e^{\s x} du \wedge dx - 2\s \sin(\s y+ h(u))e^{\s x} du \wedge dy.
\label{ex1F}
\end{equation}
It is straightforward to verify that $(g, F)$ defined by (\ref{ex1metric}) and (\ref{ex1F}) satisfy the Einstein-Maxwell equations (\ref{EM1}), (\ref{EM2}).

\subsection{Spherically symmetric and plane symmetric pure radiation spacetimes}

Here we consider a couple of classes of pure radiation solutions of the Einstein equations, obtained by assumptions of plane symmetry and spherical symmetry. These metrics depend upon a freely specifiable function $m(u)$. We use the results of \S3 and \S4 to determine whether the metrics can define null electrovacua, perhaps for some special choices of $m(u)$.  Despite considerable mathematical similarity between these two  classes of metrics, we shall see that they have very different properties with regard to the Einstein-Maxwell equations, as might be expected on physical grounds.

\subsubsection{Plane symmetry}

Let $(u, x, y, z)$, with $z > 0$, be coordinates on $M = {\bf R}^4$.  The plane symmetric, pure radiation metric is of the form \cite{Stephani}
\begin{equation}
g = \frac{2 m(u)}{z} du \otimes du - du \otimes dz - dz \otimes du + z^2(dx \otimes dx + dy \otimes dy),
\label{planemetric}
\end{equation}
where $m(u)$ is any monotonically decreasing function, $m^\prime < 0$. 
The Ricci tensor of $g$ is
\begin{equation}
{\cal R} = -\frac{2}{z^2} m^\prime(u) du \otimes du,
\end{equation}
which satisfies  all conditions in (\ref{nullRicci}).  The vector field $k$ tangent to the null congruence $\cal C$ is determined by (\ref{nullRicci2}) to be
\begin{equation}
k = \frac{2}{z} \sqrt{-2m^\prime(u)} \partial_z.
\end{equation}
A null tetrad $(k, l, m, \overline m)$ adapted to $k$ is given by
\begin{align}
&k = \frac{2}{z} \sqrt{-2m^\prime(u)} \partial_z, \quad l = \frac{1}{2\sqrt{-2m^\prime(u)}}\left(z\partial_u + m(u)\partial_z\right),\nonumber\\
& m = \frac{1}{\sqrt{2}z}\left(\partial_x + i\partial_y\right), \quad \overline m = \frac{1}{\sqrt{2}z}\left(\partial_x - i\partial_y\right)
\end{align}
The spin coefficients for this tetrad are given by
\begin{align}
&\alpha = \beta = \kappa = \lambda = \pi = \sigma = \tau = \nu = 0, \quad
\epsilon = -\frac{1}{z^2} \sqrt{-2m^\prime(u)}, \nonumber\\
&\gamma = \frac{\sqrt{2}z m^{\prime\prime}(u)}{16\left(-m^\prime(u)\right)^{3/2}},\quad
\mu = \frac{m(u)}{2z\sqrt{-2m^\prime(u)}},\quad \rho = -\frac{2}{z^2} \sqrt{-2m^\prime(u)}.
\end{align}
From $\kappa=\sigma = \Im(\rho) = 0$ we see that $\cal C$ is geodesic, shear-free, and twist-free.
Thus (\ref{planemetric}) is a null electrovacuum if and only if (\ref{twistfreeIC}) is satisfied, which is easily verified to be the case.  Thus all plane symmetric pure radiation spacetimes (\ref{planemetric}) are null electrovacua, a result already known through other methods \cite{Stephani}.  

\subsubsection{Spherical symmetry}

The spherically symmetric pure radiation spacetime is given by the Vaidya metric:
\begin{align}
g = -(1 - \frac{2m(u)}{r}) du \otimes du - &du \otimes dr - dr \otimes du\nonumber\\
& + r^2(d\theta \otimes d\theta + \sin^2\theta d\phi \otimes d\phi),
\label{sph}
\end{align}
where $r>0$, $r\neq 2m(u)$,  $\theta$ and $\phi$ are coordinates on the sphere, and where $m(u)$ is any monotonically decreasing function. 
The Ricci tensor of $g$ is 
\begin{equation}
{\cal R} = -\frac{2}{r^2} m^\prime(u) du \otimes du,
\end{equation}
which satisfies  all conditions in (\ref{nullRicci}).  The vector field $k$ tangent to the null congruence $\cal C$ is determined by (\ref{nullRicci2}) to be
\begin{equation}
k = \frac{2}{r} \sqrt{-2m^\prime(u)} \partial_r.
\end{equation}
A null tetrad $(k, l, m, \overline m)$ adapted to $k$ is given by
\begin{align}
&k = \frac{2}{r} \sqrt{-2m^\prime(u)} \partial_r, \quad l = \frac{r}{2\sqrt{-2m^\prime(u)}}\left(\partial_u - \half(1 - \frac{2m(u)}{r})\partial_r\right),\nonumber\\
& m = \frac{1}{\sqrt{2}r}\left(\partial_\theta + \frac{i}{\sin\theta}\partial_\phi\right), \quad \overline m = \frac{1}{\sqrt{2}r}\left(\partial_\theta - \frac{i}{\sin\theta}\partial_\phi\right).
\end{align}
The spin coefficients of this tetrad are
\begin{align}
&\kappa = \lambda=\nu=\sigma=\pi=\tau=0,\quad \alpha = -\frac{\sqrt{2}}{4r} \cot\theta,\quad \beta = \frac{\sqrt{2}}{4r} \cot\theta,\nonumber\\
&\epsilon = -\frac{1}{r^2}\sqrt{-2m^\prime(u)},\quad
\gamma =  \frac{1}{4[-2m^\prime(u)]^{3/2}}\left(rm^{\prime\prime}(u) + m^\prime(u)\right),\nonumber\\
&\mu = -\left(1 - \frac{2m(u)}{r}\right) \frac{1}{4\sqrt{-2m^\prime(u)}}, \quad \rho = -\frac{2}{r^2}\sqrt{-2m^\prime(u)}.
\label{ex2Spin}
\end{align}

From $\kappa=\sigma = \Im(\rho) = 0$ we see that $\cal C$ is geodesic, shear-free, and twist-free.
Thus (\ref{sph}) is a null electrovacuum if and only if (\ref{twistfreeIC}) is satisfied.  Using (\ref{ex2Spin}) we have
\begin{equation}
{\cal J}_1 = \Re\left[(\overline\delta+\overline\beta-\alpha)(\tau-2\beta)\right] - \half (\mu - \overline\mu)(\epsilon-\overline\epsilon) = \frac{1}{2r^2}.
\label{J1}
\end{equation}
Thus (\ref{twistfreeIC}) cannot be satisfied and the spherically symmetric pure radiation spacetime (\ref{sph}) can never be a null electrovacuum.  Equation (\ref{J1})  establishes the non-triviality of the invariant ${\cal J}_1$.

\subsection{Nurowski-Tafel spacetimes}

Nurowski and Tafel have constructed a class of  solutions of the Einstein-Maxwell equations \cite{Pawel} with null electromagnetic field. In particular, they have found the only known solutions which have twisting rays and a purely radiative electromagnetic field. These solutions are built from a freely specifiable holomorphic function $b(\xi)$ and two parameters, $\alpha$ and $c_2$.   We consider a special case of these solutions, specialized to Petrov type III ($c_2=0$) and with $b(\xi) = b/\xi$, where $b$ is a real constant.  To avoid confusion with the Newman-Penrose spin coefficients, we relabel $\alpha$  as the parameter $a$ in the metric.  We will verify that the integrability conditions (\ref{twistIC}) are satisfied and then construct the electromagnetic field  from the metric.

The metric can be defined by a specification of an adapted null tetrad $(k, l, m, \overline m)$. We use coordinates $(u, r, \xi, \overline\xi)$, with $\overline\xi$ being the complex conjugate of $\xi$. With the definitions
\begin{equation}
\chi := \frac{2\sqrt{2} (|\xi|^2 + 1)^2}{ \sqrt{a^2(1 - |\xi|^2)^2 + r^2(1 + |\xi|^2)^2}}
\end{equation}
\begin{equation}
\psi = \frac{1 +|\xi|^2 - |\xi|^4 + (|\xi|^2 +1)^2 \ln(|\xi|^2 +1) }{\xi^2 [a(1- |\xi|^2)- ir(1 + |\xi|^2)]},
\end{equation}
 we have for the adapted tetrad \cite{Pawel}
\begin{equation}
k = \frac{b^2}{|\xi|^3}\chi\partial_r,
\label{nt1}
\end{equation}
\begin{equation}
l = \frac{|\xi|^3}{b^2\chi}\partial_u - \frac{|\xi|}{\chi}\partial_r
\label{nt2}
\end{equation}
\begin{equation}
m =  \frac{a}{b} \psi\partial_u + \frac{2 ab\overline\xi }{a(1 - |\xi|^2) - ir(1 + |\xi|^2)}\partial_r - \frac{ib(1 + |\xi|^2)^2}{a(1 - |\xi|^2) - ir(1 + |\xi|^2)}\partial_\xi
\label{nt3}
\end{equation}
\begin{equation}
\overline m = \frac{a}{b} \overline\psi\partial_u + \frac{2 ab\xi }{a(1 - |\xi|^2) + ir(1 + |\xi|^2)} \partial_r + \frac{ib(1 + |\xi|^2)^2}{a(1 - |\xi|^2) + ir(1 + |\xi|^2)}\partial_{\overline\xi}
\label{nt4}
\end{equation}
The metric defined by this null tetrad is a pure radiation solution of the Einstein equations and satisfies (\ref{nullRicci2})

The spin coefficients for this tetrad are given by
\begin{equation}
\kappa = \lambda = \pi = \sigma = \tau = 0,
\end{equation}

\begin{equation}
\alpha = -\frac{ib[a(|\xi|^4 - 16|\xi|^2 + 3) + ir(|\xi|^4 - 2|\xi|^2 - 3)](1 + |\xi|^2)}{4\overline\xi^2[ a^2(|\xi|^2-1)^2 + r^2( |\xi|^2 + 1)^2]}
\end{equation}

\begin{equation}
\beta =  -\frac{b[a(- 3|\xi|^4 + 4|\xi|^2 + 3 ) + ir(3|\xi|^4 + 6|\xi|^2 + 3)](1 +|\xi|^2)}{\xi^2[a^2(|\xi|^2-1)^2 +r^2(|\xi|^2+1)^2]}
\end{equation}

\begin{equation}
\epsilon =  \frac{i\sqrt{2}b^2 [(2a(1 - |\xi|^2) + ir( |\xi|^2+1)](|\xi|^2+1)^3}{|\xi|^3[a^2(|\xi|^2-1)^2 +r^2(|\xi|^2+1)^2]^{3/2}}
\end{equation}

\begin{equation}
\gamma =  \frac{i\sqrt{2}|\xi|[(2a(1 - |\xi|^2) + ir( |\xi|^2+1)]}{8( |\xi|^2+1)[a^2( |\xi|^2-1)^2 +r^2(|\xi|^2+1)^2]^{1/2}}
\end{equation}

\begin{equation}
\mu =  \frac{i\sqrt{2}|\xi| [a^2(|\xi|^2-1)^2 +r^2(|\xi|^2+1)^2]^{1/2}}{4( |\xi|^2+1)[(a(|\xi|^2-1) - ir( |\xi|^2+1)]}
\end{equation}

\begin{equation}
\nu= \frac{i\xi^2[a(|\xi|^2 - 1) + ir( |\xi|^2+1)]}{8b( |\xi|^2+1)^2}
\end{equation}

\begin{equation}
\rho = -\frac{2i\sqrt{2}b^2( |\xi|^2+1)^3}{|\xi|^3[a(|\xi|^2 - 1) + ir(|\xi|^2+1)][a^2( |\xi|^2-1)^2 +r^2(|\xi|^2+1)^2]^{1/2}}.
\end{equation}

Since $\kappa = \sigma = 0$, the congruence determined by $k$ is geodesic and shear-free. Moreover,  
\begin{equation}
\omega \equiv -\Im(\rho) = \frac{2\sqrt{2} a b^2 (|\xi|^2-1) (|\xi|^2 +1)^3}{|\xi|^3[a^2(|\xi|^2-1)^2 +r^2(|\xi|^2+1)^2]^{3/2}}
\end{equation}
so the congruence is twisting.   A lengthy computation shows that the conditions (\ref{twistIC}) are satisfied.  Thus this pure radiation spacetime is indeed a null electrovacuum.  

We now construct the electromagnetic field. The equations (\ref{phieq}) and (\ref{ic0}) for the duality rotation function $\phi$ can be massaged into the form
\begin{equation}
\partial_u\phi = 0.
\end{equation}

\begin{equation}
\partial_r\phi = \frac{a (|\xi|^4 - 1)}{a^2(|\xi|^2 - 1)^2 + r^2( |\xi|^2+1)^2}
\end{equation}

\begin{equation}
\partial_\xi\phi = -{3i\over 2}\, \frac{(a^2 + r^2)(1+|\xi|^4) - 2(a^2 - r^2 + {2\over3}iar)|\xi|^2}
{a^2(|\xi|^2-1)^2 + r^2(1 + |\xi|^2)^2}
\end{equation}

\begin{equation}
\partial_{\overline\xi}\phi = {3i\over 2}\, \frac{(a^2 + r^2)(1+|\xi|^4) - 2(a^2 - r^2 - {2\over3}iar)|\xi|^2}
{ a^2(|\xi|^2-1)^2 + r^2(1 + |\xi|^2)^2}.
\end{equation}
The solution to these equations is 
\begin{equation}
\phi = {3i\over 2}\ln\left({\overline\xi\over\xi}\right) + \tan^{-1}\left({r( |\xi|^2+1)\over a(|\xi|^2-1)}\right) + const.
\end{equation}
The resulting electromagnetic field is given by
\begin{equation}
F = \cos\phi\ \Xi \wedge \Gamma - \sin\phi\ \Xi \wedge \Upsilon,
\label{NTEM}
\end{equation}
where
\begin{align}
\Xi = &-{b^2\over 2|\xi|^3} \chi\, du + i\sqrt{2} a \frac{1 +|\xi|^2 - |\xi|^4 + (|\xi|^2 +1)^2 \ln(|\xi|^2 +1) }{|\xi|^3 \xi \sqrt{a^2( |\xi|^2-1)^2 + r^2( |\xi|^2+1)^2}}\, d\xi\nonumber\\
&-i\sqrt{2} a \frac{1 +|\xi|^2 - |\xi|^4 + (|\xi|^2 +1)^2 \ln(|\xi|^2 +1) }{|\xi|^3 \overline\xi \sqrt{a^2(1 - |\xi|^2)^2 + r^2(1 + |\xi|^2)^2}}\, d\overline\xi
\end{align}
\begin{align}
\Gamma =  -{i\over \sqrt{2} b( |\xi|^2+1)^2} \Big\{&\xi[a (|\xi|^2 - 1) + ir(|\xi|^2 + 1)]d\xi\nonumber\\
& - \overline\xi[a(|\xi|^2 - 1) - ir(|\xi|^2 + 1)]d\overline\xi\Big\}
\end{align}
\begin{align}
\Upsilon =  {1\over \sqrt{2} b( |\xi|^2+1)^2} \Big\{&\xi[a (|\xi|^2 - 1) + ir(|\xi|^2 + 1)]d\xi\nonumber\\
& + \overline\xi[a(|\xi|^2 - 1) - ir(|\xi|^2 + 1)]d\overline\xi\Big\}
\end{align}
The metric determined by the null tetrad (\ref{nt1})--(\ref{nt4})  and the electromagnetic field given by (\ref{NTEM}) satisfy the Einstein-Maxwell equations.

\subsection{A twisting, pure radiation spacetime}

Let $(u, r, \xi, \overline\xi)$ be a chart on $M$, where $u$, $r$ are real and $\xi$, $\overline\xi$ are complex conjugates. The following null tetrad $(k, l, m, \overline m)$ is constructed from a twisting, pure radiation solution of the Einstein equations given in \cite{Stephani} (see equation (30.46) in \cite{Stephani} with $K=0$):
\begin{equation}
k = \sqrt{24m_0\over u} {1\over |B|}\partial_r,
\end{equation}
\begin{equation}
l=|B|\sqrt{u\over 24m_0}\partial_u + {1\over\sqrt{24m_0u} |B|} [(a_0^4 a_1^6 u^5  + a_0^2a_1^2 ur^2)|\xi|^2 + r(a_0^2a_1^4u^4 + m_0u + r^2)]\partial_r,
\end{equation}
\begin{equation}
m = {a_0a_1\overline\xi u\over \overline B}\partial_u + {2ia_0^2a_1^3u^2\overline\xi\over \overline B}\partial_r - {ia_1u\over \overline B}\partial_\xi
\end{equation}
\begin{equation}
\overline m = {a_0a_1\xi u\over  B}\partial_u - {2ia_0^2a_1^3u^2\xi\over  B}\partial_r + {ia_1u\over  B}\partial_{\overline\xi}.
\end{equation}
Here $a_0$, $a_1\neq0$, $m_0\neq0$ are real parameters and
\begin{equation}
B = a_0 a_1^2 u^2 + ir.
\end{equation}

The metric defined by this null tetrad satisfies (\ref{nullRicci2}). The spin coefficients are given by
\begin{equation}
\kappa = \sigma = \lambda=\pi=\tau=0,
\end{equation}
and
\begin{equation}
\alpha= \frac{(-3a_0a_1^2u^2 + ir)a_0a_1\xi}{4 B^2},
\quad
\beta=\frac{(-7a_0a_1^2u^2 + 3ir)a_0a_1\overline\xi}{4\overline B^2}
\end{equation}
\begin{equation}
\epsilon = - \frac{\sqrt{6m_0} r}{\sqrt{u} |B|^3},\quad \rho =- \frac{2i \sqrt{6m_0}}{\sqrt{u} B|B|}
\end{equation}
\begin{align}
\mu = &-i\sqrt{u\over 24m_0}{\overline B\over |B|^3} [(a_0^4a_1^6u^4 + 2ia_0^3 a_1^4 r u^2 - a_0^2 a_1^2 r^2)|\xi|^2\nonumber\\
 &+ i a_0^3 a_1^6 u^5 - 2a_0^2 a_1^4 r u^3 - i a_0 a_1^2 r^2 u - m_0r]
\end{align}
\begin{align}
\gamma = &- \frac{1}{8|B|^3\sqrt{6 m_0 u}}[2(a_0^4a_1^6 r u^5 + a_0^2 a_1^2 r^3 u)|\xi|^2 + 7 a_0^4 a_1^8 u^8 - 4i u^3 a_0 a_1^2 m_0 r\nonumber\\
&+ 2a_0^2 a_1^4 m_0 u^5 + 12 a_0^2 a_1^4 r^2 u^4 + 5 r^4]
\end{align}
\begin{align}
\nu =  -\frac{ia_0a_1\overline B\xi}{24m_0B^2}[a_0^3 a_1^6u^6 + 3i a_0^2 a_1^4 r u^4 - 2 a_0 a_1^2 m_0 u^3 - 3a_0a_1^2 r^2 u^2 - i r^3]
\end{align}
The null congruence is  shear-free and geodesic. The twist of the congruence is given by
\begin{equation}
\omega = - \Im(\rho) = \frac{\sqrt{24m_0} a_0 a_1^2 u^{3/2}}{|B|^3}.
\end{equation}
The twist vanishes in an open set if and only if $a_0=0$. Assuming $a_0\neq0$, the spacetime is an electrovacuum if and only if ${\cal I}_1 = {\cal I}_2 = 0$.   We have
\begin{equation}
{\cal I}_1 = \frac{6\sqrt{6m_0} a_0^3 a_1^5 \overline\xi u^{3/2}(a_0|\xi|^2 + iu)}{\overline B |B|^5} = \overline{\cal I}_2.
\label{last}
\end{equation}
  Evidently, this  pure radiation spacetime cannot be an electrovacuum if the twist is to remain non-vanishing. If the twist is set to zero via $a_0=0$ the resulting  spacetime satisfies (\ref{twistfreeIC}) and is diffeomorphic to one of the  null electrovacua of Petrov type D  originally found by Robinson and Trautman \cite{Stephani}.  The non-vanishing of the invariants ${\cal I}_1$, ${\cal I}_2$ in (\ref{last}) confirms their non-triviality.

\section{Summary}

We summarize the results of this paper by giving an algorithm which takes as input a spacetime metric and determines whether it defines a null electrovacuum and, when it does, builds the null electromagnetic field.  

\begin{itemize}

\item 
Given a metric $g$, compute its Ricci tensor $R_{ab}$.  If the Ricci tensor vanishes this is a vacuum spacetime.

\item 
 Check whether the Ricci tensor is null and positive, (\ref{nullRicci}).  If (\ref{nullRicci}) is not satisfied  the spacetime cannot be a null electrovacuum.  

\item 
From the null Ricci tensor, determine a null vector field $k^a$ by solving the quadratic equations $R_{ab} = \frac{1}{4} k_a k_b$.  

\item 
 The vector field $k^a$ determines a geodesic null congruence $\cal C$. From the first derivatives of $k^a$ check whether $\cal C$ is shear-free.  If $\cal C$ is not  shear-free, then $g$ cannot be a null electrovacuum.  

\item 
  Construct a null tetrad adapted to the shear-free, geodesic, null congruence $\cal C$, that is, a null tetrad whose first leg is $k^a$.  Compute the Newman-Penrose spin coefficients of the tetrad.  Compute the twist of $\cal C$. 

\item 
  If the twist of $\cal C$ vanishes, check (\ref{twistfreeIC}). If (\ref{twistfreeIC}) is not satisfied, $g$ does not define a null electrovacuum.  If (\ref{twistfreeIC}) is satisfied, $g$ defines a null electrovacuum and the electromagnetic field can be obtained by solving (\ref{phieq}) and then using (\ref{sdef}), (\ref{f}) and (\ref{generalF}).

\item 
  If the twist of $\cal C$ does not vanish, check (\ref{twistIC}). If these conditions are not satisfied, $g$ does not define a null electrovacuum.  If the conditions (\ref{twistIC}) are satisfied, $g$ defines a null electrovacuum and the electromagnetic field can be obtained by solving (\ref{phieq}), (\ref{ic0}) and then using (\ref{sdef}), (\ref{f}) and (\ref{generalF}).

\end{itemize}
Implementation of this algorithm requires only (a) solving algebraic equations which are at most quadratic, and (b) solving an integrable system of over-determined linear, inhomogeneous PDE of the form $d\phi = A$.

\bigskip\noindent
{\bf Acknowledgments:} I thank Ian Anderson and Pawel Nurowski for discussions. The bulk of the results given in  \S3, \S4, and \S5 were obtained using the {\sl DifferentialGeometry} package in {\sl Maple}.  This work was supported by grant OCI-1148331 from the National Science Foundation.

\appendix
\section{Appendix: Proof of Triviality of  ${\cal I}_3$, ${\cal I}_4$, ${\cal I}_5$,  ${\cal J}_2$, and ${\cal J}_3 $ }

Here we prove that the integrability conditions ${\cal I}_3={\cal I} _4 = {\cal I}_5=0$ in the twisting case, and ${\cal J}_2  = {\cal J}_3 = 0$ in the twist-free case, are always satisfied by the shear-free, geodesic, null congruence $\cal C$ determined by the null Ricci tensor. 

We will need the following results from the Newman-Penrose formalism \cite{Stewart}.  Under a spatial rotation (\ref{SR}) we have
\begin{align}
&\tau \to e^{2i\theta} \tau
\label{NPSR1}\\
&\alpha \to e^{-2i\theta}(\alpha + i\overline\delta\theta)\\
&\beta\to e^{2i\theta}(\beta + i\delta\theta)
\label{NPSR2}\\
&\epsilon \to \epsilon + i D\theta.
\label{NPSR3}
\end{align}
Under a null rotation (\ref{NR}) we have
\begin{align}
&\tau \to \tau + c\sigma + \overline c\rho + |c|^2\kappa
\label{NPNR1}\\
&\alpha \to \alpha + c\epsilon + c\rho + c^2\kappa\\
&\beta \to \beta + c\sigma +\overline c\epsilon  + |c|^2\kappa
\label{NPNR2}\\
&\epsilon \to \epsilon + c\kappa.
\label{NPNR3}
\end{align}
We also need the following Ricci identities:
\begin{align}
&D\rho - \overline\delta\kappa = \rho^2 +|\sigma|^2 + (\epsilon + \overline\epsilon)\rho -\overline\kappa \tau - (3\alpha +\overline\beta - \pi)\kappa + \Phi_{00},
\label{NPRiccia}\\
&D\mu -\delta\pi = (\overline\rho - \epsilon -\overline\epsilon)\mu + \sigma\lambda  + (\overline\pi-\overline\alpha +\beta)\pi -\nu\kappa + \Psi_2 + 2\Lambda,
\label{NPRiccih}\\
&D\alpha -\overline\delta\epsilon=(\rho + \overline\epsilon-2\epsilon)\alpha + \beta\overline\sigma - \overline\beta\epsilon - \overline\kappa\lambda + (\epsilon + \rho)\pi + \Phi_{10},
\label{NPRicci1}\\
&D\beta - \delta\epsilon = (\alpha + \pi)\sigma + (\overline\rho-\overline\epsilon)\beta - (\mu +\gamma)\kappa - (\overline\alpha -\overline\pi)\epsilon + \Psi_1
\label{NPRicci2}\\
&D\tau - \Delta\kappa = (\tau + \overline\pi)\rho +(\overline\tau + \pi)\sigma + (\epsilon - \overline\epsilon)\tau - (3\gamma + \overline\gamma)\kappa + \Psi_1 + \Phi_{01}
\label{NPRicci3}\\
&\delta\alpha - \overline\delta\beta = \mu\rho - \lambda\sigma +|\alpha|^2 + |\beta|^2 -2\alpha\beta +(\rho-\overline\rho)\gamma + (\mu - \overline\mu)\epsilon\nonumber\\
&\quad\quad\quad\quad-\Psi_2 +\Lambda +\Phi_{11},
\label{NPRiccil}\\
&\delta\rho - \overline\delta\sigma =  (\overline\alpha +\beta)\rho- (3\alpha - \overline\beta)\sigma + (\rho -\overline\rho)\tau +(\mu -\overline\mu)\kappa - \Psi_1 + \Phi_{01}.
\label{NPRicci4}
\end{align}

Henceforth we assume that the  null tetrad is adapted to the tangent to $\cal C$, so that
\begin{align}
\kappa &= \sigma = \Psi_0 = \Psi_1 =\Phi_{00} = \Phi_{01} = \Phi_{10} =  \Phi_{02} = \Phi_{20}\nonumber\\
 &=\Phi_{11}= \Phi_{12} = \Phi_{21} = \Lambda =  0
\label{SNG}
 \end{align}
 and (see (\ref{parm}))
 \begin{equation} 
\Re(\rho) = 2\Re(\epsilon).
\end{equation}

We first show that ${\cal I}_4$ and ${\cal I}_5$, defined in (\ref{trivial1}) and (\ref{trivial2}), vanish automatically in the case where the shear-free, null geodesic congruence $\cal C$ defined by the null Ricci tensor is twisting, $\omega \neq0$.   We begin by fixing a convenient class of adapted null tetrads.
From (\ref{NPSR3}), choosing $\theta\colon M\to {\bf R}$ such that
\begin{equation}
D\theta = - \Im(\epsilon),
\end{equation}
 we obtain an adapted tetrad for which 
 \begin{equation}
 \epsilon = \overline\epsilon.
 \label{gf1}
 \end{equation}  
 From (\ref{NPNR1}) and (\ref{NPNR2}), choosing $c\colon M\to {\bf C}$ such that
 \begin{equation}
  c = -{1\over i\omega}(\overline\tau -2 \overline\beta),\
\end{equation}
 we obtain a tetrad with  
 \begin{equation}
 \tau = 2\beta,
 \label{gf2}
\end{equation} 
and since $\epsilon$ is invariant under a null rotation when $\kappa = 0$, we  maintain $\epsilon=\overline\epsilon$.  Using such a tetrad,  (\ref{NPRicci2}), (\ref{NPRicci3}), (\ref{NPRicci4}), (\ref{SNG}),   (\ref{gf1}), and (\ref{gf2}) imply
\begin{equation}
(\delta + \overline\pi - \overline\alpha - \beta)\omega = 0,
\label{d}
\end{equation}
which, along with (\ref{gf1}) and (\ref{gf2}), implies that ${\cal I}_4$ and ${\cal I}_5 = \overline {\cal I}_4$ vanish with this particular choice of adapted tetrad.    As pointed out in \S4, these quantities transform homogeneously under a change in tetrad, and under $k\to-k$. Therefore they vanish for any choice of $k$ and for any choice of adapted tetrad.

To prove that ${\cal I}_3$ given in (\ref{IC3}) vanishes  we again fix a tetrad satisfying $\epsilon = \overline\epsilon$ and $\tau = 2\beta$.  With such a tetrad
\begin{equation}
{\cal I}_3 =  \omega^2\left[(\Delta - 2\Re(\gamma))\omega - (D + 2\epsilon)\Im(\mu)\right].
\label{I3gauge}
\end{equation}
Now return to (\ref{d}) which implies
\begin{equation}
[\bar\delta, \delta](\omega) = \overline\delta[(\overline\alpha + \beta - \overline\pi) \omega)]- \delta[(\alpha + \overline\beta - \pi) \omega)].
\label{br}
\end{equation}
Use (\ref{brackets}) to evaluate the bracket on the left side of (\ref{br}). In the resulting identity, use (\ref{NPRiccia}), (\ref{NPRiccih}), and (\ref{NPRiccil}), along with (\ref{gf1}) and (\ref{gf2}), to find  that (\ref{I3gauge}) vanishes for the tetrad being used.  The transformation properties of the integrability conditions now guarantee that ${\cal I}_3$ vanishes for any tetrad adapted to $k$ and for any choice of $k$.

In the twist-free case, $  \omega = -\Im(\rho)=0$, we can still fix a tetrad with $\epsilon = \overline\epsilon$, and now we distinguish two cases: (a)  $\rho=2\epsilon =0$ and (b) $\rho=2\epsilon \neq0$.  In case (a) (\ref{NPRicci1}) and (\ref{NPRicci2}) imply $D\alpha = D\beta = 0$ and it follows that ${\cal J}_2 = 0 = {\cal J}_3$.  In case (b), a null rotation with 
\begin{equation}
c = - {1\over\rho}(\alpha - \overline\beta)
\end{equation} sets 
\begin{equation}
\alpha = \overline\beta,
\end{equation} 
still maintaining $\epsilon = \overline\epsilon$.  Equations (\ref{NPRicci1}) and (\ref{NPRicci2}) now imply that $\pi = 0$, whence (\ref{NPRicci3}) implies $(D-\overline\rho)\tau = 0$.  Eqs. (\ref{NPRicci2}) and (\ref{NPRicci4}) now imply that $(D - \overline\rho)\beta = 0$, whence ${\cal J}_2 = 0 = {\cal J}_3$ for this particular adapted tetrad.  Since these quantities transform homogeneously under a change in tetrad, and under $k\to-k$, they vanish for any choice of $k$ and for any choice of adapted tetrad.
\end{document}